%% file: ppm.tex
\documentclass[%
 preprint,
 amsmath,amssymb,
prb,
]{revtex4-1}

\usepackage{lineno,hyperref}
\usepackage[normalem]{ulem}
\modulolinenumbers[5]
\usepackage{color}

\usepackage{xspace}
\usepackage{setspace}
\usepackage{natbib}

\usepackage{graphicx}

\newcommand{\Cred}[1]{%
  {\color{black}{#1}}\xspace}

\newcommand{\beginsupplement}{%
        \setcounter{table}{0}
        \renewcommand{\thetable}{S\arabic{table}}%
        \setcounter{figure}{0}
        \renewcommand{\thefigure}{S\arabic{figure}}%
     }

\DeclareMathOperator*{\argmin}{arg\,min} 

\begin{document}

\title{Point-Pattern Matching Technique for Local Structural Analysis in Condensed Matter}



\author{Arash D. Banadaki}
\affiliation{ Department of Materials Science and Engineering, North Carolina State University, Raleigh, NC}

\author{Jason J. Maldonis}
\affiliation{Department of Materials Science and Engineering, University of Wisconsin-Madison, Madison, WI}

\author{Paul M. Voyles}
\affiliation{Department of Materials Science and Engineering, University of Wisconsin-Madison, Madison, WI}

\author{Srikanth Patala}%
\email{Corresponding Author: spatala@ncsu.edu}
\affiliation{ Department of Materials Science and Engineering, North Carolina State University, Raleigh, NC}


\begin{abstract}

  The local arrangement of atoms is one of the most important predictors of
  mechanical and functional properties of materials. However, algorithms for
  identifying the geometrical arrangements of atoms in complex materials systems
  are lacking. To address this challenge, we present a point-pattern matching
  algorithm that can detect instances of a “template” structure in a given set
  of atom coordinates. To our knowledge this is the first geometrical comparison
  technique for atomistic configurations with very different number of atoms,
  and when the optimal rotations and translations required to align the
  configurations are unknown. The pattern matching algorithm can be combined
  with an appropriate set of metrics to quantify the similarity or dissimilarity
  between configurations of atoms. We demonstrate the unique capabilities of the
  point-pattern matching algorithm using two examples where the automated
  analysis of local atomistic structure can help advance the understanding of
  structure-property relationships in material science: (a) identifying local
  three-dimensional polyhedral units along interfacial defects, and (b) the
  analysis of quasi-icosahedral topologies in the atomistic structure of
  metallic glasses. We anticipate that the pattern matching algorithm will be
  applicable in the analysis of atomistic structures in broad areas of condensed
  matter systems, including biological molecules, polymers and disordered
  metallic systems. An online implementation of the algorithm is provided via
  the open source code hosted on GitHub
  (\href{https://github.com/spatala/ppm3d}{https://github.com/spatala/ppm3d}).

\end{abstract}

\flushbottom
\maketitle

\thispagestyle{empty}


\newpage

\input{Introduction}

\input{Methodology}

\input{Results}

\input{Conclusions}

\clearpage
\newpage

\bibliography{ppm}

\clearpage
\newpage

\input{supp_info}

\end{document}

%% file: Introduction.tex
\section{Introduction}
\label{sec:Intro}

In materials science, structure is typically described using atoms as
fundamental units and the properties are inferred through the spatial
arrangements of atoms relative to each other. The length scales involved may
vary from short-range (near-neighbors) and medium-range to long-range depending
on the structure and properties of interest. Even when the analysis of structure
at larger length scales is necessary, the characterization of the relative
arrangement of atoms in the first coordination shell has proven to be of great
importance \cite{ding2014soft, lazar2015topological}. For perfect crystals, the
machinery of crystallography provides a complete, succinct, and extremely
powerful description of the positions of all the atoms, but there is no parallel
description for aperiodic structures, including defects in crystalline materials
such as grain boundaries and the structure of glasses. In such cases, we do not
have a general description (or quantification) of even the local structure. A
parallel problem arises in atomistic simulations. In this case, the position of
every atom is known, often as a function of time, yielding a great deal of data,
but leaving us with a need for a general, automated analysis approach to develop
abstract structural descriptions from this detailed data.

Some of the most commonly utilized structural analysis methods for atomistic
simulations include: centro-symmetry parameter, bond-order analysis,
common-neighbor analysis, bond angle analysis, and Voronoi-cell topology. The
first four techniques involve computing for each atom in the system a scalar
quantity, an order-parameter, which depends on the spatial location of the
atom's neighbors. Voronoi-cell topology uses a topological descriptor, the
Voronoi index or the $p$-vector \cite{barnette1969p, lazar2012complete}, instead
of a scalar calculated from spatial locations as the order parameter. These
quantities are invariant to rigid body rotations and translations of the local
atomic environment. While these simplified order parameters have been invaluable
in the analysis of the distributions of different types of defects and their
evolution during atomistic simulations, they suffer from issues involved with
degeneracies and large sensitivity to small perturbations. The degeneracies
arise as the order parameters are simply projections of a high-dimensional
atomistic configuration space to a single scalar value, as described in Lazar
\textit{et al}. \cite{lazar2015topological} For example, the centro-symmetry
parameter exhibits similar values for either an atom that is present along a
defect (such as a dislocation or a grain boundary) or one that is in a bulk
single crystal at moderately high temperatures (\textit{e.g.} 0.5 $T_m$).

The issue of high sensitivity to small perturbations is evident when using the
Voronoi cell topology as a descriptor. For any topological descriptor, there
will exist atomic configurations where a small change in the position of one of
the atoms will change the topology of the descriptor
\cite{lazar2015topological}. Another limitation of the structure classifiers is
that all the quantities measured are per-atom descriptors. That is, each atom
gets a scalar value or a topological index. However, in complex systems,
\textit{e.g.} along defects in crystalline materials (such as dislocations and
grain boundaries) or in quantifying medium-range order in glasses \cite{}, it is
the geometrical patterns and the connectivity of a sub-set of atoms in the
system that are of interest. \Cred{In these systems there is no clear notion of
  a \emph{center-atom} around which the structure is quantified. Therefore,
  traditional structure quantification techniques are not capable of discerning
  the connectivity between atoms in complex, disordered material systems.}

In order to address these challenges, we propose a point-pattern matching
technique for direct characterization of local atomic structures. Point-pattern
matching (PPM) is a fundamental problem in pattern recognition with applications
in a broad range of fields including computer vision \cite{chang1997fast,
  myers2000bayesian, carcassoni2003spectral, wang2004kernel,
  goodrich1999approximate, caetano2009learning}, computational chemistry
\cite{martin1993fast, finn1997rapid}, astronomy \cite{murtagh1992new,
  weber1994application} and computational biology \cite{akutsu2003point,
  nussinov1991efficient}. More specifically, PPM methods are \textit{feature
  detection algorithms} that can identify specific features in an environment.
PPM techniques use a metric to quantify the similarity (or dissimilarity)
between atomic environments, as described in section \ref{sec:met}. This allows
for a quantitative measure of how structure changes, for example, as temperature
is increased or when the crystallographic nature of the defect is varied (see
section \ref{sec:3dpu}). PPM algorithms can also account for small perturbations
that arise due to thermal vibrations, making them ideal for the analysis of
local atomic structure in disordered material systems. Most notably, with a
suitable metric, unsupervised machine learning algorithms, such as clustering,
can be used to analyze the underlying geometries in the material
\cite{MEinprep}.

In the following sections, we first introduce the PPM algorithm with a simple
two-dimensional example. Then, we present two material science problems
illustrating the unique capabilities of the PPM algorithm for analyzing
structures in atomistic models. The first is the identification of polyhedral
motifs in the disordered regions of a metallic grain boundaries \Cred{(section
  \ref{sec:3dpu}). In this example, the objective is to identify model
  polyhedral units \cite{banadaki2017three} along the atomic structure of grain
  boundaries (GBs). The polyhedral units are much smaller than the GB structure,
  so this application takes advantage of the capability of PPM to match sets
  with very different number of atoms. In the second example (section
  \ref{sec:glass}), we use the PPM technique to identify clusters with
  icosahedral geometry in a Zr-Cu-Al metallic glass for comparison to clusters
  with icosahedral topology as identified by their Voronoi indices.}

%% file: Methodology.tex
\section{Methodology}
\label{sec:met}

In the terminology used in image-processing, atoms are referred to as points and
the sets of atoms being compared are point-sets. To quantify the (dis)similarity
between two point-sets, we compare a set of $m$ points (the \textit{model},
$\mathbf{M}$, also known as the \textit{template}) to a set of $n$ points (the
\textit{target}, $\mathbf{T}$). The positions of the points in the model,
$\mathbf{M}$, and in the target, $\mathbf{T}$, are given by the set of vectors
$\mathbf{R^M} = \{ \mathbf{r}_1^M, \mathbf{r}_2^M, \ldots, \mathbf{r}_m^M \}$
and $\mathbf{R^T} = \{ \mathbf{r}_1^T, \mathbf{r}_2^T, \ldots, \mathbf{r}_n^T
\}$, respectively. The objective is to find a set of $m$ points in the target
that are most similar to the configuration of points in the model set. That is,
we wish to find a one-to-one mapping between the atoms in the model and the
target, a rotation matrix, and a translation vector that best overlaps the
points in the model with those in the target set.

In the general problem, the number of points $n$ in the target is greater than
those in the model, \textit{i.e.} $n \geq m$. A brute-force technique requires
picking ${m \choose n}$ sub-sets of $m$ points in the target, and finding the
correspondence requires another $m!$ comparisons. Hence, the algorithm, in the
worst case, has the complexity of $O\left(n!/(n-m)!\right)$, which is
computationally unfeasible for large values of $n$ and $m$. A wide variety of
approaches have been proposed in the fields of image processing
\cite{salvi2007review, goshtasby2012image} and computational chemistry
\cite{coutsias2004using, bartok2010gaussian, sadeghi2013metrics,
  ferre2015permutation, behler2016perspective, de2016comparing,
  griffiths2017optimal, temelso2017arbalign, imbalzano2018automatic,
  musil2018machine} to render this general problem computationally feasible .

\Cred{In condensed-matter systems, the goal is to compare and characterize
  molecular structures. When the number of atoms in the model and the target are
  equal (\textit{i.e.} $n = m$),} the similarity (or dissimilarity) between
point-sets is quantified using the root-mean-square-distance (RMSD) parameter,
defined as \cite{sadeghi2013metrics}:

\begin{equation}
    \text{RMSD} = \frac{1}{\sqrt{m}} \min_{f ,U, \mathbf{t}} \sum_{i=1}^m
    d \left( U \mathbf{r}_i^M + \mathbf{t}, \mathbf{r}_{f(i)}^T \right)\text{,}
    \label{eq:rmsd}
\end{equation}

where $d(\cdot, \cdot)$ is the Euclidean distance metric. The minimization is
over all possible permutations of indices, given by the function $f$, rotations
of the model, $U$, and relative translations $\mathbf{t}$. The permutation
function $f$ maps the indices of the points in the model to those in the target.
For example, if $f(i^m) = j^t$, then the $i^{th}$ index in the model is mapped
to the $j^{th}$ index in the target (the same information may also be
represented using a $m \times m$ permutation matrix). We denote the operations
$f$, $U$ and $\mathbf{t}$ that minimize RMSD as $\hat{f}$, $\hat{U}$, and
$\mathbf{\hat{t}}$, respectively. The RMSD defined as such obeys the three
properties of a metric \cite{rudin1976principles}: (a) the coincidence axiom,
(b) symmetry, and (c) the triangle inequality.

If the minimizing rotations and translations, $\hat{U}$, and $\mathbf{\hat{t}}$,
are known \emph{a priori}, the permutation matrix can be obtained using the
Hungarian algorithm \cite{kuhn1955hungarian}, which has the algorithmic
complexity of $O(m^3)$. If only the permutation matrix, $\hat{f}$, is known
\emph{a priori}, the parameters $\hat{U}$ and $\mathbf{\hat{t}}$ can be computed
using Horn's algorithm \cite{horn1987closed} with complexity $O(m)$. Horn's
algorithm uses quanternions to provide a closed-form solution for aligning two
point sets such that RMSD is minimized. If all of $\hat{f}$, $\hat{U}$, and
$\mathbf{\hat{t}}$ are unknown, more sophisticated algorithms are required for
minimizing RMSD between clusters of atoms. A good review of different techniques
is provided in Ref. \cite{griffiths2017optimal}. For example, in Ref.
\cite{sadeghi2013metrics}, a two-stage method, which includes a Monte-Carlo
perturbation, has been proposed for finding the global-minimum in RMSD.
\Cred{Another recent technique uses molecular dynamics and simulated annealing
  \cite{fang2010atomistic, fang2011spatially} to align clusters, and has been
  used to identify first neighbor (short-)\cite{fang2010atomistic} and first-
  through third-neighbor (medium-) range order \cite{fang2011spatially} in
  metallic glasses}. In Ref. \cite{griffiths2017optimal}, two efficient
algorithms, GO-PERMDIST and FASTOVERLAP, were introduced. However, these
techniques currently are limited to equal number of atoms in the model and the
target. And, more importantly, the optimal translation $\mathbf{\hat{t}}$ that
produces the best match is assumed to be known \emph{a priori}. While this is a
reasonable approximation when $n = m$, determining the translational component
is non-trivial when the two structures to be compared have different number of
atoms.

\subsection{An Illustration of the Two-Dimensional Point-Pattern Matching Algorithm}
\label{sec:ppm_supp}

Here we present a new Point-Pattern Matching (PPM) algorithm based on graph
theory arguments developed by McAuley and Caetano \cite{mcauley2012fast}. This
technique is general enough to be applied for systems with different number of
atoms. While Ref. \cite{mcauley2012fast} contains complete details of the
rigid-graph PPM algorithm, in this section, we provide an illustration of the
steps in the algorithm for point-sets in two-dimensions.

\begin{figure}[h!]
  \centering
  \includegraphics[width=0.9\linewidth]{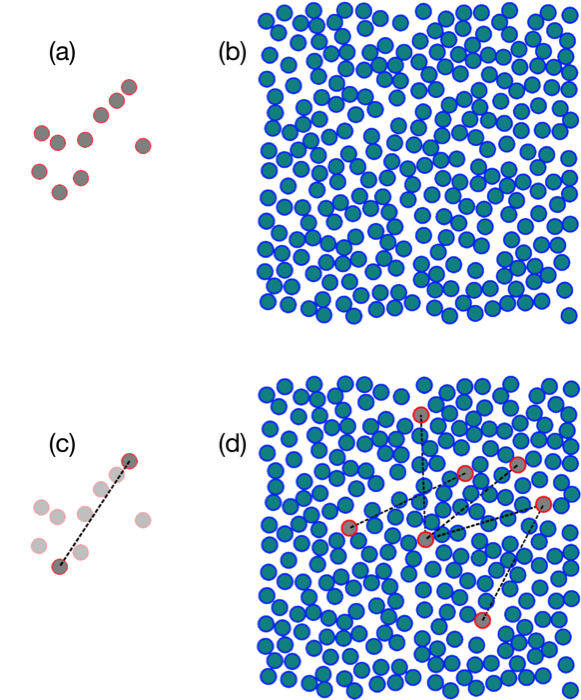}
  \caption{The model and target point-sets are shown in (a) and (b),
    respectively. The points are represented using discs to illustrate
    atoms in this example. In (c), the two root nodes in the model are
    highlighted. In (d), a few example pairs of points in the target
    that match the root nodes are illustrated.}
  \label{fig:ppmv1}
\end{figure}

Figure \ref{fig:ppmv1}(a, b) show the model and the target, respectively. The
objective is to identify the set of points in the target that best match the
model point-set. In Ref. \cite{mcauley2012fast}, a graph representation of a
point-set is used to identify the optimal permutation mapping $\hat{f}$. The
sets of points in the model $\mathbf{M}$ and target $\mathbf{T}$ can be
expressed as graphs by using the information contained in the atom positions
(nodes) and bonds (edges). Usually, a graph is defined by the set of nodes and
edges $(V, E)$. We will, however, consider a graph $\mathcal{G}$ to be a set of
pairs of nodes, and say $(i,j) \in \mathcal{G}$ if an edge connecting nodes $i$
and $j$ is present in graph $\mathcal{G}$ (using the notation developed in Ref.
\cite{mcauley2012fast}). The complete graphs, which contain all the possible
edges in the point-set, representing the model and the target will be denoted by
$\mathcal{M}$ and $\mathcal{T}$, respectively.

Instead of using RMSD defined in Eq. \ref{eq:rmsd}, McAuley and Caetano
introduced a metric that depends only on edge lengths and hence is invariant to
rotations and translations. This metric, for a given permutation function $f$,
is defined on the graphs and is provided in Eq. \ref{eq:edgd}. The objective of
the PPM technique is to determine the function $\hat{f}$ that minimizes the
metric $\mathcal{D}(f)$ as defined in Eq. \ref{eq:fullg}.

\begin{equation}
    \mathcal{D}(f) = \sum_{(i,j) \in \mathcal{M}}  \left| d \left( \mathbf{r}_i^M, \mathbf{r}_j^M \right) - d \left( \mathbf{r}_{f(i)}^T, \mathbf{r}_{f(j)}^T \right)\right|
    \label{eq:edgd}
\end{equation}

\begin{equation}
  \hat{f} = \argmin_{f: \mathcal{M} \rightarrow  \mathcal{T}} \mathcal{D}(f)
  \label{eq:fullg}
\end{equation}

\begin{figure}[h!]
    \centering
    \includegraphics[width=\linewidth]{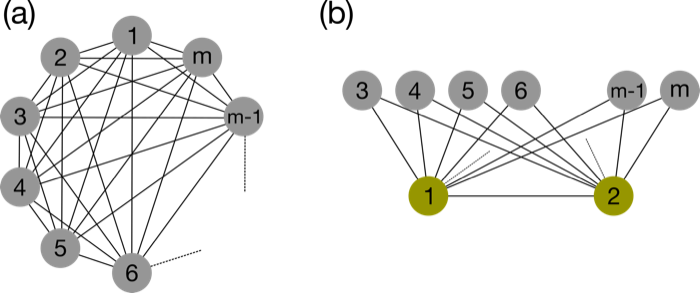}
    \caption{(a) A complete graph with all possible edges connecting the $m$ 
    points in the model is shown. In (b), the rigid-graph of the model is shown. 
    The rigid graph in two dimensions consists of the edge between the
    two root nodes (highlighted) and all the edges between the rest of points 
    and the root nodes.}
    \label{fig:rnodes}
\end{figure}

The mapping $\hat{f}$ produces a matching between the points in the model and
the target, such that the sum of the differences in all the edge lengths is
minimized. This definition allows for computing optimal mappings even in the
presence of noise. The algorithms to determine $\hat{f}$ fall under the class of
quadratic assignment problems, which are in general NP-hard
\cite{anstreicher2003recent}. In Ref \cite{mcauley2008graph}, McAuley
\textit{et al.} introduced the concept of a rigid-graph to solve the matching
problem in an efficient manner. A rigid-graph is a subset of the original graph
such that the only transformations that can be applied to the node coordinates
while preserving the distances in the rigid-graph are isometries (rigid-body
translations and rotations). If the rigid-graph of the model $\mathcal{M}$ is
denoted by $\mathcal{R}$, then the new objective function can be written as:

\begin{equation}
  \hat{f} = \argmin_{f: \mathcal{R} \rightarrow  \mathcal{T}} \mathcal{D}(f) \quad \text{where}  \quad \mathcal{D}(f) = \sum_{(i,j) \in \mathcal{R}}  \left| d \left( \mathbf{r}_i^M, \mathbf{r}_j^M \right) - d \left( \mathbf{r}_{f(i)}^T, \mathbf{r}_{f(j)}^T \right)\right|
  \label{eq:rigg}
\end{equation}
where $\mathcal{M}$ is replaced by the rigid-model-graph $\mathcal{R}$. The
difference between $\mathcal{M}$ and $\mathcal{R}$ is the number of edges
considered during the minimization. For example, a complete graph, where all the
possible edges in the model point-set are considered to be a part of the graph,
is shown in Figure \ref{fig:rnodes} (a). In this complete graph, there are
$m(m-1)/2 = O(m^2)$ edges. As defined in Ref.~\cite{mcauley2012fast}, the rigid
graph for a two-dimensional (2D) point-set is constructed by first choosing two
nodes that are connected by an edge. These two points are usually referred to as
the \emph{root nodes}. All the other points in the model are then connected to
the two root nodes as shown in Figure \ref{fig:rnodes} (b). Therefore, the rigid
graph of the model contains the edges between the root nodes and, instead of all
the edges in the model point-set, only those edges that connect the rest of the
points to the root nodes. The total number of edges in this rigid graph is $2m-3
= O(m)$. The global rigidity theorem \cite{caetano2006graphical} implies that
preserving the lengths of the edges in $\mathcal{R}$ preserves the lengths in
the complete graph of the model point-set. The required number of root nodes
depends on the dimensionality of the point-set. In 2D, two root nodes are
required and, in 3D, we need three root nodes to construct the rigid graph.


\begin{enumerate}

\item The input for the PPM algorithm contains the two point-sets and the root
  nodes in the model. The choice of the root nodes does not influence the
  accuracy of the algorithm but if the root nodes are chosen judiciously, the
  run-time for obtaining the solution can be reduced. We are interested in
  minimizing the sum of differences in the edge lengths in the rigid graph of
  the model and the mapped point-set in the target (Eq. \ref{eq:rigg}). We
  assume that these differences scale as the actual edge length. Therefore,
  picking the nodes that contain the largest edge as the root nodes can improve
  the efficiency of the algorithm. With this assumption, we choose the two nodes
  that are farthest apart as root nodes in 2D. The root nodes for the model are
  shown in Figure \ref{fig:ppmv1}(c). In 3D, we choose three nodes where the sum
  of the edge lengths of the triangle is maximized.

\item Once the root nodes are fixed, we find all possible matches in the target
  for the root nodes. In 2D, the computational cost is $O(n^2)$. This is
  equivalent to enumerating all possible pairs of points in the target. In 3D
  the cost is $O(n^3)$ as three root nodes are required. Some of the matches for
  the root nodes in the target are shown in Figure \ref{fig:ppmv1}(d). Consider
  the best match for the root nodes, as shown in Figure \ref{fig:ppmv2}(a, b). A
  local coordinate system is defined with the root nodes in the model (Figure
  \ref{fig:ppmv2}(a)) and the matched-root-nodes in the target (Figure
  \ref{fig:ppmv2}(b)).

\item For each remaining point in the model, the vector in the local coordinate
  system is computed (as shown for two points in the model in Figure
  \ref{fig:ppmv2}(c)). An equivalent vector in the target is then defined using
  the local coordinate system in the target (Figure \ref{fig:ppmv2}(d)). A
  nearest-neighbor algorithm is then used to determine the point $n_p$ that is
  closest to the the equivalent vector in the target. For the two vectors in the
  model, equivalent vectors and the nearest points are shown in the target in
  Figure \ref{fig:ppmv2}(d). The algorithmic complexity to determine the nearest
  point is $O(\log n)$ (using a KD-tree data-structure).

\item Finally, the steps described above are repeated for all the remaining
  points in the model (Figure \ref{fig:ppmv3}(a)). The points in the target that
  best match the model are shown in Figure \ref{fig:ppmv3}(b).

\item In Figure \ref{fig:ppmv3}(c), the alignment between the model and the
  mapped points are shown by overlapping points. To align the two point
  patterns, the rotation $U$ and rigid body translation $\mathbf{t}$ that
  minimizes the RMSD metric defined in Eq. \ref{eq:rmsd} is computed using
  Horn's algorithm \cite{horn1987closed}, which has a complexity of $O(m)$.


\begin{figure}[h!]
  \centering
  \includegraphics[width=0.9\linewidth]{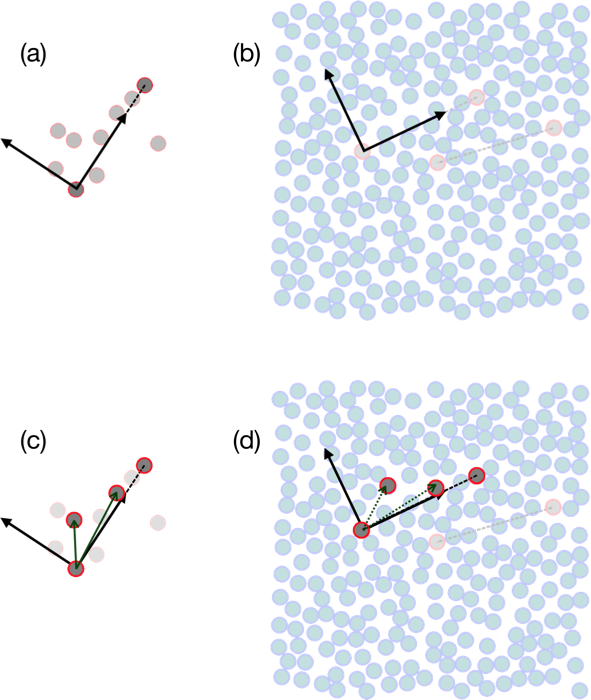}
  \caption{The local coordinate systems, defined by the root nodes, are
    illustrated in the model and the target in (a) and (b), respectively. In (c)
    the vectors of two points (other than the root nodes) in the model are shown
    and in (d) the equivalent vectors are shown using dotted lines. The
    highlighted points in (d) are the nearest-points to the equivalent vectors
    in the target.}
  \label{fig:ppmv2}
\end{figure}


  \begin{figure}[h!]
    \centering
    \includegraphics[width=\linewidth]{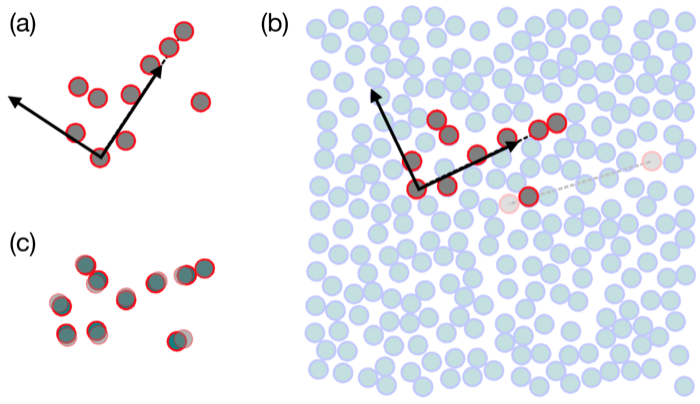}
    \caption{The model point-set with the local coordinate system is shown in
      (a). The best match found in the target point-set is shown in (b). In (c),
      the registration between the model and the mapped-points in the target,
      which is obtained using Horn's algorithm, is shown.}
    \label{fig:ppmv3}
  \end{figure}

\end{enumerate}

The complexity of the point pattern matching algorithm to compute $\hat{f}$ is
determined by combining the complexities in steps 1-3 and is given by $O(m n^d
\log n)$, where $d=2,3, \ldots$ is the dimensionality of the point-set. The PPM
algorithm is made reasonably efficient by using a rigid-graph representation (so
as to reduce the total number of edge-lengths to be matched), and by using the
$k$-nearest neighbor algorithm for identifying similar points in the model and
the target. The Horn's algorithm in the last step adds an additional
computational cost of $O(m)$ but the dominant terms come from the PPM algorithm.



\begin{figure}[h!]
\centering
\includegraphics[width=0.6\linewidth]{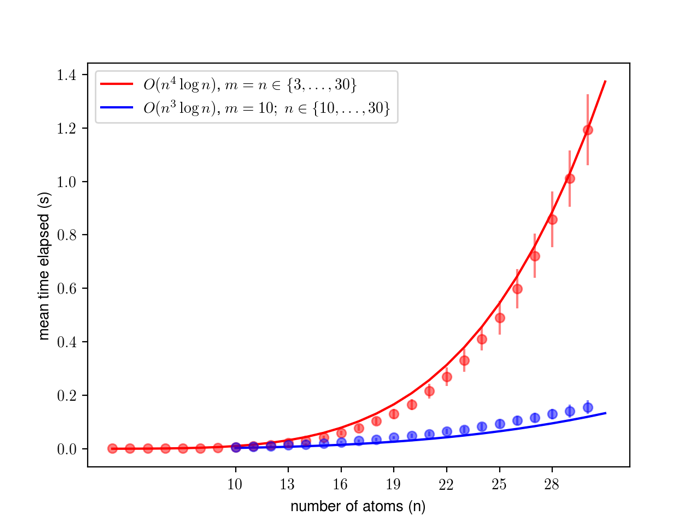}
\caption{Peformance scaling for the PPM algorithm as a function of the number of
  points in the model and target sets are shown. The tests are performed for
  equal number of points in both the model and target, i.e. $n=m$ and are shown
  in the red curve. The scaling with a fixed number of points in the model set
  $(m=10)$ and increasing number of points in the target is shown in the blue
  curve. The point sets are three-dimensional and the scaling is $O(mn^3
  \log(n))$.}
\label{fig:bigO}
\end{figure}

In Figure \ref{fig:bigO}, we illustrate the complexity of the algorithm by
comparing clusters of equal and unequal number of atoms, respectively. For the
red curve, there are equal number of points in the model and target (i.e.
$n=m$), and for the blue curve the number points in the model is fixed $(m =
10)$ and the number of points in the target are varied. The target point-sets
are created by picking random points in three-dimensions with the $x$, $y$ and
$z$ coordinates in the interval $[-1, 1]$. The model set is created by picking a
random subset of the target, adding a Gaussian random noise (with $\sigma =
0.1$) to the points, and randomizing the indices. The alignments are performed
on a standard desktop computer (Intel Xeon quad-core processor, 1.80 GHz, 4 GB
RAM) and the asymptotic scaling of $O(m n^3 \log n)$ for three-dimensional
point-sets is shown in Figure \ref{fig:bigO}. The PPM algorithm is coded in C++,
with a Python wrapper to promote ease of use, and is shared online at
\href{https://github.com/spatala/ppm3d}{https://github.com/spatala/ppm3d}. The
python wrapper has been implemented with capabilities for parallelization on a
distributed computing platform, such as HTCondor \cite{thain2005distributed},
for applications where millions of alignments have to be performed.

\subsection{Potential Applications}
\label{sec:apps}

\Cred{PPM is particularly useful when the number of atoms in the clusters being
  compared is not the same and the translation vector $\mathbf{\hat{t}}$ is not
  known. Several important problems in condensed matter systems, both within
  materials science and in other fields, fall in this category. A few examples
  problems are discussed here, then results for two of them are presented in
  detail in the next section.

  Section \ref{sec:3dpu} discusses the problem of comparing grain boundary (GB)
  structures. To develop structure-property relationships for crystalline
  interfaces, it is necessary to investigate structural variations as function
  of the crystallographic parameters \cite{han2017grain}. Banadaki and Patala
  have developed the void-clustering algorithm to create a topological
  representation of a GB structure using polyhedral units
  \cite{banadaki2017three}. However, the polyhedral unit representation is a
  topological descriptor and, as is the case with any topological descriptor,
  small changes in the atom positions can completely alter the polyhedral units
  observed in the GB. Thus, it is not robust against small perturbations in the
  crystallographic parameters of the GB, nor is it a continuous function of
  those parameters. These limitations make it impossible to use void-clustering
  and the polyhedral representation to compare GB structures to one another in a
  meaningful way. The PPM algorithm removes these limitations. As described in
  section \ref{sec:3dpu}, we use the void-clustering algorithm to define the
  \emph{model} polyhedral units in \emph{singular} GBs. Then, we use PPM to find
  these model units in other GBs that are \emph{vicinal} to the singular GBs. In
  this problem, the model structure (the polyhedral unit) and the target (the
  entire vicinal GB) have very different numbers of atoms, and the optimal
  translations, required to identify the polyhedral units, are unknown.

  Similar methodology could be used to identify near-coincident site lattices
  (CSLs) in special GBs.\cite{patala2017approximating}. CSLs are matching
  lattice-sites between two crystals. These lattices can be either different
  phases or lattices with different orientations. The CSLs determine orientation
  relationships during phase transformations and are used as one of the criteria
  for determining the crystallography of low-energy interfaces. To date, there
  does not exist a general algorithm to determine all the possible CSLs (or
  near-CSLs) in a given material system. PPM could be used to identify the
  coincidence sites by enumerating sublattices and comparing them to each other.
  In this problem, the number of atoms in the model sublattice is significantly
  smaller than the number of atoms in the target phase, making this a good
  potential problem for PPM.

  Section \ref{sec:glass} discusses the problem of characterizing local
  structures in metallic glasses. Metallic glasses are often represented as
  characteristic first-neighbor atomic clusters which pack together to make a
  solid \cite{sheng2006atomic}. Those clusters are identified by their
  polyhedral shape, which for metal-metal glasses like Zr-Cu is often
  icosahedral \cite{DingPRB2008CuZrandPdCuSi, mareview, ShengZCAPotential}.
  However, the inherent structural disorder in a glass means that only a small
  fraction of the polyhedra are perfectly icosahedral. Others are identified as
  ``quasi''-icosahedral \cite{sheng2006atomic}, often based on the idea that
  they are fundamentally an icosahedron, but with an extra atom or a missing
  atom due to disorder. We use PPM to test whether or not clusters with (quasi-)
  icosahedral topologies have icosahedral \emph{geometry} as well. In this case,
  the model is a geometrically perfect icosahedron consisting of twelve atoms,
  and the targets are first-neighbor clusters drawn from a metallic glass model
  with coordination numbers varying from 9 to 16. PPM both provides the ability
  to match such clusters to one another and to define a quantitative and
  continuous similarity metric for comparison.

  Beyond nearest-neighbor clusters, PPM could be used to characterize
  larger-scale medium-range order in metallic glasses in two ways. First, one
  could identify all the SRO clusters conforming to a particular geometry
  (\emph{e.g.} an icosahedron), then identify if they are connected by, for
  example, how many atoms the connected clusters have in common. This approach
  is commonly used with topological structure measures. Second, PPM can be used
  to identify clusters with atoms outside the first-neighbor shell, provided
  that one can define a model geometry. Candidate model geometries could include
  particular connected geometries of icosahedra (e.g. a face-sharing pair)
  \cite{sheng2006atomic} or larger Bergman or Mackay clusters
  \cite{ClusterAlignmentOnZCA}. As the number of atoms in the model cluster
  increases, the ability to match a perfect model cluster to target clusters
  with varying coordination number in order to accommodate structural disorder
  becomes even more important.

  Beyond the topics discussed here, PPM could be useful in modeling the growth
  of nanoclusters and mutations in biomolecules. In the area of nanoparticle
  synthesis and stability, different pathways along which nanoparticles grow is
  of interest \cite{wales2000energy}. To construct these pathways using
  atomistic simulations, one has to compare and align clusters of different
  sizes. These comparisons are also necessary for calculating free energies and
  rates of different pathways. Mutation of biomolecules has a parallel
  ``pathways'' problem, as we often wish to track changes as a function of
  generation and calculate free energies and rates along the
  way.\cite{wales2006energy} Since the number of atoms change during the growth
  and mutation processes, PPM is well-suited to this task.
    
  Finally, the PPM algorithm can also be utilized when the extent of overlap
  between the model and the target point-sets is incomplete. That is, when there
  are outliers in the model point-set that do not match with points in the
  target. Such outliers are referred to as \emph{occlusions}
  \cite{sonka2014image} in the image-processing literature. When occlusions are
  allowed, the PPM algorithm finds the largest subset of the model and the
  target that results in the best possible matching. This capability is
  particularly useful when there is a limited overlap between the clusters being
  compared, \emph{e.g.} when identifying binding regions between
  proteins.\cite{comin2009binding, padhorny2016protein}
    
}

%% file: Results.tex
\section{Results and Discussion}
\label{sec:Results}

\input{Results_GBs}

\input{Results_Glasses}

%% file: Results_GBs.tex
\subsection{Local Atomic Motifs in Grain Boundaries -
  A Three-Dimensional Polyhedral Unit Model}
\label{sec:3dpu}

GBs are planar defects in polycrystalline materials and influence a wide array
of structural and functional properties \cite{sutton1995interfaces}.
Unfortunately, GBs are also one of the least understood defect types in
materials science. This is due to the vast and topologically complex,
five-dimensional crystallographic degrees-of-freedom (DOF) of a GB
\cite{patala2012improved,patala2013symmetries,homer2015grain}. The five
dimensions correspond to the so-called macroscopic degrees of freedom - three
parameters define the misorientation between individual grains and the other two
fix the boundary-plane orientation. These parameters constitute the
bicrystallographic aspects of interfaces. One of the primary objectives of grain
boundary engineering has been to predict structure and properties of GBs as a
function of the five crystallographic parameters.

From a geometrical perspective, the structure of certain GBs (those with
low-index planes at the interface) has traditionally been represented using
clusters of atoms that form quasi-two-dimensional geometrical motifs. This
model, referred to as the structural unit model (SUM), was first proposed by
Bishop and Chalmers \cite{bishop1968coincidence} and has been extended to a
variety of tilt GBs by Sutton and Vitek \cite{sutton1983structurev1,
  sutton1983structure}. More recently, Han et al. developed a framework
utilizing the metastable-SUM to predict GB structures and energies for $[100]$
and $[111]$ symmetric-tilt GBs\cite{han2017grain}. They emphasized that the
metastable-SUM framework can be used to describe structural variations in the
complete five-dimensional crystallographic phase-space of GBs. One of the key
steps in this framework relies on \emph{identifying pre-determined geometrical
  motifs in a variety of minimum-energy and metastable GB structures.}
Identifying three-dimensional geometrical motifs in complex, disordered GBs is
difficult, but the PPM algorithm is uniquely suited to address this issue.

\begin{figure}[h!]
  \centering
  \includegraphics[width=\linewidth]{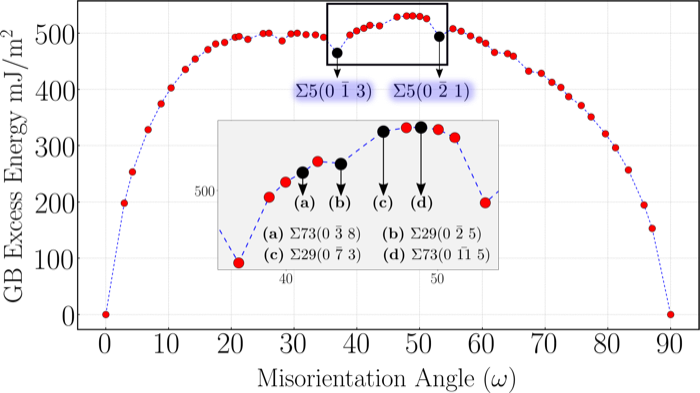}
  \caption{The energy of grain boundaries in aluminum as a function of
    misorientation angle. The energies of the $[100]$ symmetric tilt GBs and the
    other GBs analyzed in detail ((a) - (d)) are marked with arrows and labeled
    with their crystallographic elements.}
  \label{fig:stilt_eng}
\end{figure}

Rather than focusing on the physical aspects of GB properties, here we show a
simple example to illustrate how the PPM algorithm can be used to describe
structural variations in GBs as a function of crystallographic parameters such
as misorientation angle. \Cred{We will use the polyhedral
  units\cite{ashby1978structure} to compare GB structures, instead of the
  somewhat arbitrarily defined structural units. We can use the void-clustering
  algorithm \cite{banadaki2017three}, which is based on the clustering of voids
  present in the GB structure, to automate the process of representing a GB
  structure as a combination of polyhedral units. We then want to use this
  representation to compare the atomistic structures of different GBs with
  similar crystallography. As mentioned previously, the void-clustering
  algorithm by itself cannot reach this goal. Instead, we adopt a two-step
  procedure in which void-clustering is first used to identify the model
  polyhedral units in the \emph{singular} GBs. Singular GBs correspond to cusps
  in the energy landscape \cite{bulatov2014grain}, so their structure is less
  sensitive to perturbations. Then, we use the PPM technique to identify the
  polyhedral units found in the singular boundaries (the model point-sets) in
  the vicinal GB structures (the target point-sets). Quantifying the density and
  the spatial arrangement of the model polyhedral units will provide a direct
  link between the properties of the singular and the vicinal GBs
  \cite{balluffi1983simple, han2017grain}. To illustrate this two-step process,
  we analyze the set of $[100]$ symmetric tilt GBs in aluminum with the
  misorientation angle ranging from $36.87^{\circ}$ to $53.13^{\circ}$. Figure
  \ref{fig:stilt_eng} shows the energies of the $[100]$ symmetric tilt GBs as a
  function of the tilt angle.}

\begin{figure}[p]
  \centering
  \includegraphics[width=0.8\linewidth]{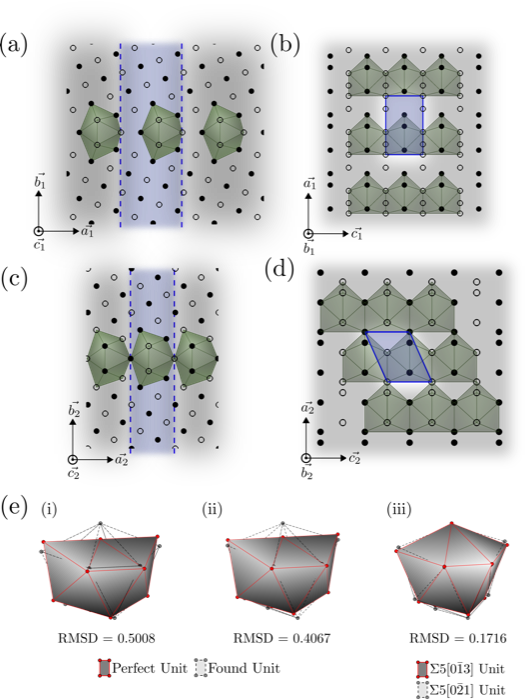}
  \caption{The atomistic structure of $\Sigma 5 \left( 0 \, \bar{1} \, 3
    \right)$ and $\Sigma 5 \left( 0 \, \bar{2} \, 1 \right)$ with the 11-atom
    polyhedral units highlighted. The views in (a, c) are along the tilt axis,
    $[1 0 0]$, and (b, d) are along the boundary-plane normal. In (e) the RMSD
    of the observed polyhedral units compared to the perfect octadecahedra and
    compared to each other is reported. The ``Found Unit'' in the legend of (e,
    i) and (e, ii) refers to the unit observed in the $\Sigma 5 \left( 0 \,
      \bar{2} \, 1 \right)$ and $\Sigma 5 \left( 0 \, \bar{1} \, 3 \right)$ GBs,
    respectively. The axes in the (a, b) are such that, $\vec{a}_1 = [0 \, 3 \,
    1]$, $\vec{b}_1 = [0\, \bar{1}\, 3]$ and $\vec{c}_1 = [1 0 0]$. The axes in
    (c, d) are, $\vec{a}_2 = [0 \, 1 \, 2]$, $\vec{b}_2 = [0\, \bar{2}\, 1]$ and
    $\vec{c}_2 = [1 0 0]$.}
  \label{fig:s5_gbs}
\end{figure}

\Cred{The first step is to identify the polyhedral units in the singular GBs. In
  Figure \ref{fig:stilt_eng}, the two singular $\Sigma 5$ GBs, corresponding to
  cusps in the energy landscape, are highlighted. Banadaki and Patala showed
  that 11-atom polyhedra, which can be viewed as distorted octadecahedra
  \footnote{This polyhedron is more precisely referred to as the
    Edge-Contracted-Icosahedron (ECI) \cite{wiki:eci}. In chemistry, however,
    the ECI is most commonly called the octadecahedron, for 18 triangular
    faces.}, are observed in these two $\Sigma 5$ GBs.\cite{banadaki2017three}.
  Figure \ref{fig:s5_gbs} shows their atomistic structures and polyhedral unit
  representations.}

\clearpage
\newpage

\clearpage
\newpage

\begin{figure}[h!]
  \centering
  \includegraphics[width=\linewidth]{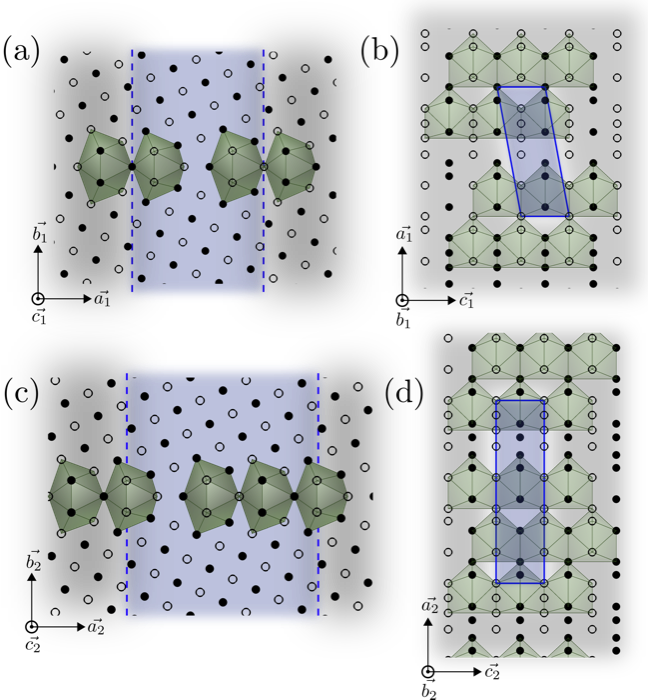}  
  \caption{The atomistic structure of $\Sigma 29 \left( 0 \,\bar{2} \, 5
    \right)$ and $\Sigma 29 \left( 0 \, \bar{7} 3 \right)$, with the 11-atom
    polyhedral units highlighted, in (a, b) and (c, d). The views in (a, c) are
    along the tilt axis and (b, d) are along the boundary-plane normal. The axes
    in the (a, b) are such that, $\vec{a}_1 = [0 \, 5 \, 2]$, $\vec{b}_1 = [0\,
    \bar{2}\, 5]$ and $\vec{c}_1 = [1 0 0]$. The axes in (c, d) are, $\vec{a}_2
    = [0 \, 3 \, 7]$, $\vec{b}_2 = [0\, \bar{7}\, 3]$ and $\vec{c}_2 = [1 0
    0]$.}
  \label{fig:s29_gbs}
\end{figure}

\Cred{The second step is to use the PPM algorithm to compare the structures of
  the two singular $\Sigma 5$ GBs with the vicinal GBs ($\Sigma 29$ and $\Sigma
  73$ GBs, also highlighted in Figure \ref{fig:stilt_eng}). That is, we want to
  identify the octadecahedral units (the model point-sets, containing 11 atoms)
  in the vicinal GB structures (the target point-sets, containing many atoms).
  For example, in the $\Sigma 73 \left( 0 \, \bar{11} \, 5 \right)$ GB, there
  are about 265 atoms in the GB structure. The optimal translation vector
  $\hat{\mathbf{t}}$ is also unknown.

  The octadecahedral units identified using the PPM algorithm in the $\Sigma 29
  \left( 0 \,\bar{2} \, 5 \right)$ and $\Sigma 29 \left( 0 \, \bar{7} 3 \right)$
  GBs are shown in Figure \ref{fig:s29_gbs}. The octadecahedra found in the
  structures of $\Sigma 73 \left( 0 \, \bar{3} \, 8 \right)$ and $\Sigma 73
  \left(0\, \overline{11} 5 \right)$ GBs are shown in Figures \ref{fig:s73_gb1}
  and \ref{fig:s73_gb2}, respectively, in the Supplementary Information. These
  results show that the structures of the $\Sigma 73$ and $\Sigma 29$ vicinal
  GBs can be expressed as a combination of the octadecahedral units observed in
  the singular $\Sigma 5$ GBs. The variations in the tilt angle are accommodated
  by changing the spatial arrangement of the octadecahedral units and by adding
  ``gaps'' between the units. These gaps correspond to the $\mathbf{B}$ or
  $\mathbf{I}$ structural units from the perspective of the SUM
  \cite{priedeman2018quantifying} or the dual-tetrahedra and octahedra from the
  perspective of the polyhedral unit model \cite{banadaki2017three}. }

The physical significance of comparing GB structures like this has been
extensively discussed in the literature.\cite{han2017grain,
  sutton1983structurev1, sutton1983structure, balluffi1983simple,
  balluffi1996should}. For example, Han \emph{et al.} showed that interfacial
energies can be predicted for the $[100]$ and $[111]$ symmetric tilt GBs in BCC
tungsten over the entire misorientation range based on atomistic simulations of
only four delimiting, singular GBs,\cite{han2017grain} and Balluffi has proposed
a simple model for predicting properties, such as diffusivity, of symmetric tilt
GBs using the structural unit model and the properties of singular
interfaces.\cite{balluffi1983simple}, \Cred{The ability to express the structure
  of vicinal GBs as combinations of polyhedral units that are characteristic of
  particular low-energy, singular GBs, therefore, allows us a straightforward
  way to understand and predict their properties. The combination of
  void-clustering and PPM described here represents a powerful tool for
  developing quantitative structure-property relationships for complex GBs,
  \emph{i.e.} interfaces with mixed crystallographic character in the complete
  five-dimensional phase-space.}

\clearpage
\newpage

%% file: Results_Glasses.tex
\subsection{Local Structure in Metallic Glasses - The Quasi-Icosahedral
  Clusters}
\label{sec:glass}

Metallic glasses (MGs) are disordered materials, and their lack of long-range
translational symmetry necessitates a rigorous understanding of their short- and
medium-range order (SRO and MRO) structure. SRO in MGs is comprised of an atom
and its nearest neighbors, which we have been referring to as a cluster. SRO
balances efficient packing on one hand and chemical ordering on the other
\cite{laws2015predictive}. Efficient packing of clusters without long-range
translational symmetry requires the structure of the material to be disrupted,
and therefore identifying the structural units that cause this disruption and
lack of connectivity at the short-range length scale is of considerable
interest. For example, in Zr-Cu based MGs icosahedra with 5-fold symmetry (which
cannot tile 3D space) are the dominant SRO structural motifs
\cite{sheng2006atomic, DingPRB2008CuZrandPdCuSi, mareview, ShengZCAPotential}.
In addition, the distribution of clusters such as icosahedra can have a profound
influence on the mechanical properties of bulk metallic glasses
\cite{ding2014soft}. For further information, a review on MG structure can be
found in Ref \cite{mareview}.

Characterization of clusters in MGs is done most often using Voronoi indices,
the $p$-vector of the number of three-, four-, five-, and six-sided faces. For
example, in Zr-Cu based glasses, the atoms with icosahedral SRO are identified
as those with Voronoi indices $\langle 0 \; 0 \; 12 \; 0 \rangle$. However, the
clusters will not have a perfect icosahedral geometry due to the inherent
disorder in the glass. In addition, previous studies classified some Voronoi
indices with a high number of pentagonal faces and geometry intuitively related
to an icosahedron as \emph{quasi-}icosahedral \cite{sheng2006atomic}, even if
they contain 11 or 13 atoms. However, while the quasi-icosahedral Voronoi
polyhedra resemble an icosahedron, a quantitative basis for this consideration
is lacking due to the topological nature of the Voronoi method.

Here, we employ the PPM algorithm to analyze the atomic structure of a
Zr$_{50}$Cu$_{45}$Al$_5$ model glass. \Cred{Similar glasses have been studied
  using other alignment methods. For example, Fang \textit{et. al}
  \cite{ClusterAlignmentOnZCA} studied a Cu-Zr MG using their atomic cluster
  alignment method \cite{ClusterAlignmentMethod} and identified SRO and MRO
  structures in the glass. The cluster alignment method uses a collective
  alignment scheme to align a set of hundreds to thousands of clusters
  simultaneously and collectively. The collective alignment is excellent at
  identifying the overarching structure of a set of clusters, but is not adept
  at quantifying the similarity or difference of pairs of clusters as the PPM
  algorithm does. Here we use PPM to address the following two questions about
  the Zr$_{50}$Cu$_{45}$Al$_5$ model glass}:

\begin{enumerate}
\item How distorted are the clusters with Voronoi topology $\langle 0 \; 0 \; 12
  \; 0 \rangle$, when compared to a geometrically perfect icosahedron? and
\item How similar are the clusters whose Voronoi polyhedra have a high number of
  5-sided faces (some of which are considered quasi-icosahedral) to a perfect
  icosahedron?
\end{enumerate}

\begin{figure}
  \centering
  \includegraphics[width=\linewidth]{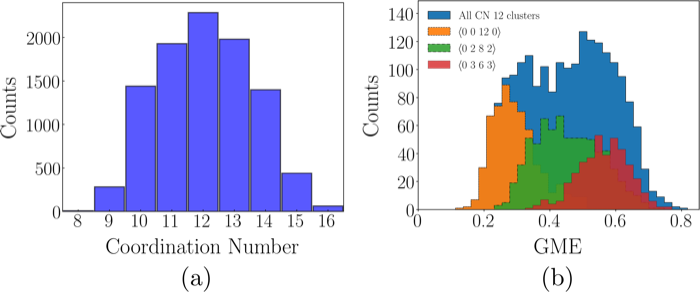}
  \caption{(a) The coordination number distribution for the
    Zr$_{50}$Cu$_{45}$Al$_5$ metallic glass model studied in this section. (b)
    The bimodal distribution (blue) of GMEs for all 2,285 clusters with
    coordination number 12 in the MG model. The colors show analogous histograms
    for the sets of clusters with three different Voronoi indices, all of which
    have coordination number 12.}
\label{fig:cn_distribution+vps_histogram}
\end{figure}

The Zr$_{50}$Cu$_{45}$Al$_5$ model used in this work was obtained by quenching a
liquid with that composition with 9,826 atoms from 2000 K to 600 K at $5 \times
10^{10}$ K/s using molecular dynamics in LAMMPS \cite{LAMMPS} using the Sheng
embedded atom model potential \cite{ShengZCAPotential}, updated in 2012. After
quenching, the glass was equilibrated for $500$ ps and the inherent structure
was calculated by performing a conjugate gradient minimization of the potential
energy. We extracted every cluster from this model. The Voronoi index
distribution of these clusters is identical to those of other models produced by
the same potential \cite{ShengZCAPotential}. The coordination number
distribution of the clusters is shown in Figure
\ref{fig:cn_distribution+vps_histogram}(a). Coordination number twelve is the
most common, consistent with icosahedral SRO. For each cluster, the bond lengths
from the center atom to its nearest neighbors were normalized so their average
value was 1.0, then each cluster was compared to a perfect icosahedron using
PPM. \Cred{Normalizing bonds lengths before matching with PPM is not required by
  the algorithm, and we have tested the method both with and without
  normalization. In this case, our goal is to test whether icosahedral topology
  is a good predictor of icosahedral geometry, so we elected to normalize the
  bond lengths before PPM, which provides a better match to topological measures
  of icosahedra. For example, normalizing the bond lengths makes alignment
  against a perfect icosahedron less sensitive to the composition of the
  clusters. Without normalization, PPM would report a better match to an
  icosahedron for a Zr-rich, Al-centered cluster with longer but uniform
  nearest-neighbor distances than for a cluster closer to the mean composition
  of the glass. In cases where the bond-lengths / relative atomic radii are
  important, differences in the radii can be easily recover, as they the
  normalization factors are recorded by the code. Horn’s algorithm also
  calculates the optimal scaling factor when comparing clusters of different
  sizes (once the mapping between the indices is obtained using the PPM
  algorithm).}

After alignment, three metrics in addition to the objective function, RMSD, were
calculated for each cluster comparing it to the perfect icosahedron: $L^1$,
$L^{inf}$, and a measure of angular variance, $V_A$. In this section we refer to
RMSD as $L^2$ to highlight the connection between RMSD, $L^1$, and $L^{inf}$.
These metrics are defined by:

\begin{equation}
  \begin{split}
    &L^1(\mathbf{M}, \mathbf{T}) = \frac{1}{m}\sum_{i=1}^{m} | \mathbf{\hat{r}}_i^M - \mathbf{r}_{f(i)}^T | \\
    &L^{inf}(\mathbf{M}, \mathbf{T}) = \max(\mathbf{\hat{r}}_i^M - \mathbf{r}_{f(i)}^T) \\
    &V_A(\mathbf{M}, \mathbf{T}, c) = \frac{1}{N}\sum_{n=i,j}^{N} | \angle (\mathbf{\hat{r}}_i^M, \mathbf{0}, \mathbf{\hat{r}}_j^M) - \angle (\mathbf{r}_{f(i)}^T, \mathbf{0}, \mathbf{r}_{f(j)}^T) |
  \end{split}
\end{equation}
\noindent where $m$ is the number of points in the model and target,
$\mathbf{\hat{r}}_i^M = U \mathbf{r}_i^M + \mathbf{t}$ is the position of point
$i$ in the model point-set after rotation and translation, and the $max$ in
$L^{inf}$ runs over the indices $i \in [1, m]$. $N$ is the number of neighbors
(bonds) in the model where two points are neighbors if they are within a
distance $c$ of one another, $\angle(\cdot, \mathbf{0}, \cdot)$ is the angle
between a pair of points going through the center of the point-set, and the
summation over $n=i,j$ in $V_A$ includes all pairs of neighbors in the
point-set. In this work we define neighbors using the cutoff $c = 3.6$ \AA,
which is the first minimum in the \Cred{total} radial distribution function.

These four metrics were chosen to quantify various aspects of the differences in
atomic structure as well as to illustrate the ability to calculate various
structural similarity metrics after alignment by PPM. $L^1$ puts less emphasis
on outliers \Cred{(poorly matching atoms in the model and the target)} than
$L^2$, while $L^{inf}$ only considers the worst outlier. $V_A$ provides a
measure of angular variation to complement the bond length measures of the $L^p$
metrics. The geometric mean of these four metrics, $\sqrt[4]{L^2 \cdot L^1 \cdot
  L^{inf} \cdot V_A}$, was used as the final metric of comparison to the perfect
icosahedron and is henceforth called the geometric mean error (GME). \Cred{The
  GME of all the metrics was found to better separate different structures than
  any single metric.} The geometric mean is appropriate for calculating averages
of numbers with different numerical ranges and retains information about the
relative change of those numbers when comparing different values.

We first consider all 2,285 clusters with coordination number 12 after alignment
to a target of a perfect icosahedron. Figure
\ref{fig:cn_distribution+vps_histogram}(b) (blue) shows the distribution of the
GMEs calculated after these alignments. The distribution is bimodal, and the
low-GME peak is composed of geometrically icosahedral clusters. GME histograms
for clusters with Voronoi indices $\langle0$ $0$ $12$ $0\rangle$, $\langle0$ $2$
$8$ $2\rangle$, or $\langle0$ $3$ $6$ $3\rangle$ (orange, green, and red,
respectively) illustrate the range of GME values associated with clusters with
these different topologies.

\begin{figure}
    \centering
    \includegraphics[width=\linewidth]{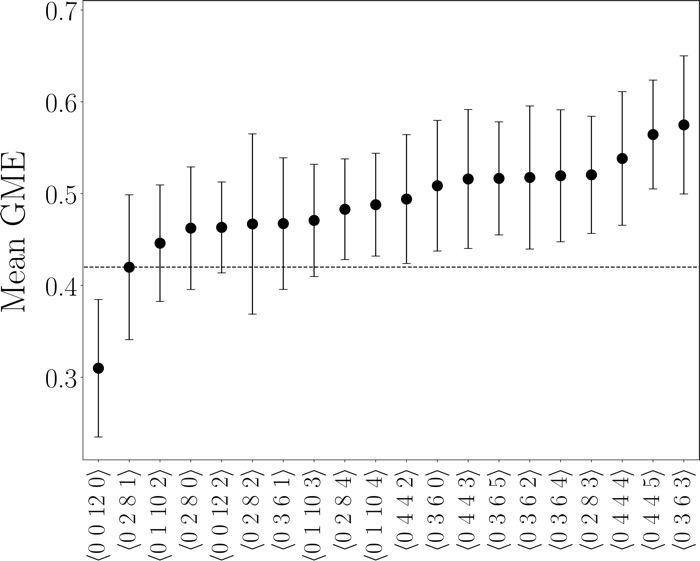}
    \caption{The mean of the GME for clusters with specific Voronoi indices with
      a high number of pentagonal faces. The clusters with Voronoi index
      $\langle 0 \; 0 \; 12 \; 0 \rangle$ are well below the cutoff of $0.42$,
      while clusters with Voronoi index $\langle 0 \; 3 \; 6 \; 3 \rangle$ are
      well above the cutoff and are therefore do not have icosahedral geometry
      despite their high number of pentagonal faces and coordination number of
      12. The error bars show one standard deviation of the GME values. The
      horizontal dotted line designates the GME cutoff of $0.42$.}
    \label{fig:mean_GME_of_VP_categories}
\end{figure}

\Cred{Next we consider all of the clusters in the model, regardless of
  coordination number, but still aligned against the icosahedron.} Figure
\ref{fig:mean_GME_of_VP_categories} shows the mean and standard deviations of
the GME for clusters with the most common topologies, categorized by their
Voronoi indices. The clusters with Voronoi indices $\langle0 \; 0 \; 12 \;
0\rangle$ are most geometrically similar to the perfect icosahedron, consistent
with the MG literature, as shown by their notably low mean GME. The Voronoi
indices that are most often considered quasi-icosahedral include $\langle0 \; 2
\; 8 \; 2\rangle$, $\langle0 \; 2 \; 8 \; 1\rangle$, and $\langle0 \; 1 \; 10 \;
2\rangle$ \cite{mareview, WangRSCAtomisticStudy, CrystalGenes,
  StructureOfZ73Pt27, GuoFeDopedZCA}, and while many of the clusters with these
topologies exhibit a low GME, they display a wide range of distortions. In
addition, the mean GME of clusters with Voronoi index $\langle 0 \; 3 \; 6 \; 3
\rangle$ is $0.57$. This value is significantly larger than the dip between the
two peaks in the total histogram in Figure
\ref{fig:cn_distribution+vps_histogram}(b) at $0.42$, so these clusters should
not be classified as quasi-icosahedral based on their geometry. Clusters with VI
$\langle0 \; 0 \; 12 \; 0\rangle$ or $\langle0 \; 2 \; 8 \; 2\rangle$ and GME
greater than $0.42$ tend to be Zr-centered ($83\%$ and $62\%$, respectively).
This indicates that Zr-centered clusters with icosahedral topology tend to be
more distorted than Cu- or Al- centered clusters with icosahedral topology,
consistent with previous findings \cite{ShengZCAPotential}. The average
composition of the nearest-neighbor shells of these same clusters is similar to
the overall composition of the model, so there are no compositional
abnormalities in the shells of these clusters.
\begin{figure}
    \centering
    \includegraphics[width=\linewidth]{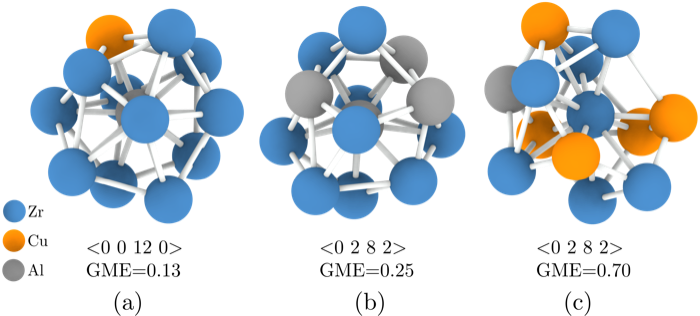}
    \caption{Three clusters illustrate the topological insufficiency of Voronoi
      indices to differentiate the geometry of the structures, while the GME
      provides sufficient descriptive power. A $\langle0$ $0$ $12$ $0\rangle$
      cluster (a) with low GME has similar structure to a $\langle0$ $2$ $8$
      $2\rangle$ cluster (b) with low GME, but different structure than another
      $\langle0$ $2$ $8$ $2\rangle$ cluster (c) with high GME.}
    \label{fig:vp1}
\end{figure}

The distribution of GME for $\langle0$ $2$ $8$ $2\rangle$ topology clusters
(green in Figure \ref{fig:cn_distribution+vps_histogram}(b)) straddles this
$0.42$ dividing line between icosahedral and non-icosahedral geometries, despite
being widely considered quasi-icosahedral in the MG literature \cite{mareview,
  WangRSCAtomisticStudy, CrystalGenes, StructureOfZ73Pt27, GuoFeDopedZCA}.
Figure \ref{fig:vp1} illustrates the geometrical disparity between two
$\langle0$ $2$ $8$ $2\rangle$ clusters. The $\langle0$ $2$ $8$ $2\rangle$
cluster in Figure \ref{fig:vp1}(b) has a low GME of $0.25$ and has similar
structure to both the perfect icosahedron and the Figure \ref{fig:vp1}(a)
cluster with $\langle0$ $0$ $12$ $0\rangle$ topology and GME $0.13$ (Figure
\ref{fig:vp1}(a)). However, the $\langle0$ $2$ $8$ $2\rangle$ cluster in Figure
\ref{fig:vp1}(c) has a large GME of $0.70$ and is dissimilar in structure to
both previous clusters.

These results answer the two questions above. $\langle0 \; 0 \; 12 \; 0\rangle$
topology is strongly associated with icosahedral geoemtry, as shown by the
histogram in orange in Figure \ref{fig:cn_distribution+vps_histogram}(b), almost
all of which falls in the low-GME peak of the total distribution below the
$0.42$ cutoff. As illustrated in Figure \ref{fig:mean_GME_of_VP_categories},
many of the $\langle 0 \; 2 \; 8 \; 1 \rangle$ topology clusters are
geometrically similar to a perfect icosahedron, as are many clusters with
$\langle 0 \; 1 \; 10 \; 2 \rangle$ and $\langle 0 \; 2 \; 8 \; 2 \rangle$
topologies. However, some clusters with the latter Voronoi indices have a GME
that falls above the geometrically-icosahedral cutoff of $0.42$ and should not
be classified as having icosahedral geometry; Figure \ref{fig:vp1}(c) shows a
specific example. $\langle0$ $3$ $6$ $3\rangle$ topologies, despite having a
fairly large fraction of 5-sided faces and being sometimes considered
quasi-icosahedral in the literature \cite{CrystalGenes, GuoFeDopedZCA} do not
have icosahedral geometry based on their GME histogram. In general, PPM
alignment and the GME score provides a quantitative measure of how icosahedral a
cluster is in a way that Voronoi indices as a measure of topology do not. PPM
and GME is a particularly useful discriminator for topologies reported to
correspond to distorted, quasi-icosahedral structures.

The question of whether geometry or topology is more important for determining
the influence of structure on the properties and processes of metallic glasses
remains to be answered. Icosahedral topology is considered important because
plastic deformation tends to avoid regions in the structure with a high
concentration of icosahedral topology \cite{PengPlasticDeformationMG};
icosahedral topology regions in the supercooled liquid have slower local
dynamics than other topologies \cite{MaPropensityCuZrPRB,
  JakseStructureinCuZrLiquid}; and the concentration of icosahedral topology
increases significantly as the liquid cools through the glass transition
\cite{MaPropensityCuZrPRB, JakseStructureinCuZrLiquid, mareview}. However, the
energy of clusters depends more on their geometry (bond lengths, bond angles,
coordination numbers) than on their topology, so we speculate that the
explanatory power of topology arises because it is a proxy for geometry that is
robust against disorder and easy to compute. PPM provides a robust, computable
method of assessing geometry; future work will test its explanatory power for
properties and processes in metallic glass systems.

%% file: Conclusions.tex
\section{Conclusions}

We have presented a point-pattern matching algorithm for local structural
analysis in atomistic simulations. The PPM algorithm relies on matching the
edges in the model point-set with those in the target. The complexity of the
algorithm for matching point sets in three-dimensions is $O \left( m n^3 \log n
\right)$, where $m$ and $n$ are the number of points in the model and the
target, respectively. While there exist efficient algorithms for matching sets
of atoms \cite{griffiths2017optimal}, these techniques generally assume that the
optimal translation required to minimize RMSD is known \emph{a priori}. However,
when $\mathbf{\hat{t}}$ is unknown, the PPM algorithm is necessary to determine
the optimal matching between the model and target point sets. \Cred{PPM is
  capable of matching structures which do not comprise the same number of
  atoms.}

\Cred{Two examples illustrated the capabilities of the PPM algorithm. First, we
  analyzed the atomic structures of grain boundaries vicinal to the two singular
  $\Sigma 5$ GBs in aluminum. The \emph{model} point-set is the 11-atom
  octadecahedral unit identified in the $\Sigma 5$ GBs using the void-clustering
  algorithm. The target point set is the entire atomic structure of the vicinal
  GBs. The PPM technique identified structures similar to the octadecahedron in
  the vicinal $\Sigma 29$ and $\Sigma 73$ GBs. When identifying polyhedral units
  in the vicinal GBs, there is no clear notion of a ``center'' atom that can be
  used as reference and, hence, the optimal translation $\hat{t}$ is unknown.
  The number of atoms in the polyhedral unit (the model) and the GBs (the
  target) are also very different. Using PPM in this way enables the extension
  of models for predicting grain boundary properties from special structural
  units to more general boundaries.}
    
\Cred{Second, we illustrated the ``hypothesis testing'' application of PPM to
  discern the presence of quasi-icosahedral topologies in the atomistic
  structure of metallic glasses. Voronoi indices are the most common tool used
  to analyze short-range order in glasses. While both PPM and Voronoi indexing
  produce a structure descriptor, PPM provides a geometric descriptor, while
  Voronoi indexing produces a topological descriptor. Topological techniques are
  sensitive to small changes in the atomic structure, which can result in
  drastic, unintuitive changes in the descriptor. For the PPM geometric
  descriptor, small changes in the atomic structure always result in small
  changes in the geometric mean error descriptor. As a result, the GME of
  PPM-aligned clusters is a meaningful, continuous ``structural distance''
  between two atomic structures. Both icosahedral and non-icosahedral clusters
  with varying coordination number were matched to a perfect icosahedron. While
  all of the clusters with perfect icosahedral Voronoi indices were also
  geometrically icosahedral, only some of the clusters with quasi-icosahedral
  geometry were geometrically icosahedral. Since geometry is more strongly
  connected to interatomic forces and energies than topology, we speculate that
  some of the quasi-icosahedral topology clusters with non-icosahedral geometry
  may have different influence the properties of the glass than their
  geometrically icosahedral counterparts.}


More broadly, PPM has the potential to find applications in the study of a
variety of materials-science and chemistry phenomena. For example, in grain
boundary science and engineering, PPM could be used to identify near-coincident
site lattices and their corresponding $\Sigma$-misorientations, which play a
fundamental role in the analysis of interfaces both in experiments and
simulations. Near-CSLs are particularly useful for identifying preferred
orientation relationships between dissimilar materials \cite{zhang2017near}. In
metallic glasses, PPM could be used to test the hypotheses that other,
non-icosahedral structures with non-crystallographic symmetry like tri-capped
trigonal prisms are present in glass-forming alloys that do not exhibit
icosahedra \cite{GaskellTrigonalPrisms}, and more generally to investigate the
role of geometry as opposed to topology on processes like the glass transition
and plastic deformation and properties like ductility. Beyond the systems
studied here, PPM can be used to systematically identify the changes in
atomistic structures when modeling the growth of nano-clusters or mutations in
polymeric/protein molecules \cite{wales2000energy, ferrando2008nanoalloys,
  schon2001determination, schon2001determinationb}.

As mentioned in section \ref{sec:met}, another unique capability of the PPM
algorithm is that one can allow for outliers (or occlusions) in the model
point-set. This capability will be particularly useful when one is not aware of
the appropriate model unit. When occlusions are allowed, the PPM algorithm finds
the largest subset of the model and the target that gives the best possible
matching. When combined with an unsupervised learning algorithm, this capability
can help identify atomic motifs \cite{MEinprep} that are common across large
clusters of atoms.

Finally, it is of interest to improve the computational efficiency of the
algorithm while being able to find the global-minima in the pattern matching.
For example, a recently proposed technique, termed Go-ICP \cite{yang2016goicp}
might provide similar results with better scaling with the size of the target.
There is, however, a computational overhead as the optimization is performed in
the space of rotations and translations (denoted by the group $SE(3)$).
Therefore, we anticipate that for smaller cluster sizes, the PPM algorithm will
more appropriate. An open-source implementation of PPM suitable for atomistic
materials simulations and parallelized computation is hosted on GitHub at
\href{https://github.com/spatala/ppm3d}{https://github.com/spatala/ppm3d}.


\section*{Acknowledgements}

Development of the PPM approach and application to grain boundaries by AB and SP
was supported by the Air Force Office of Scientific Research Young Investigator
Program funded through the Aerospace Materials for Extreme Environments
(Contract \# FA9550-17-1-0145). Computing resources for generating grain
boundary structures was provided by the High Performance Computing Center at
North Carolina State University. Implementation of the python wrapper and
parallelization and application to metallic glasses by JJM and PMV was supported
by NSF DMR-1332851, then by NSF DMR-1728933. The computing for parts of this
research was performed using the compute resources and assistance of the
UW-Madison Center for High Throughput Computing (CHTC) in the Department of
Computer Sciences. The CHTC is supported by UW-Madison, the Advanced Computing
Initiative, the Wisconsin Alumni Research Foundation, the Wisconsin Institutes
for Discovery, and the National Science Foundation, and is an active member of
the Open Science Grid, which is supported by the National Science Foundation and
the U.S. Department of Energy's Office of Science.

\newpage


%% file: supp_info.tex
\clearpage

\newpage

\beginsupplement

\begin{center}
  \textbf{ \LARGE Supplemental Materials: \\ Point-Pattern Matching
    Technique for Local Structural Analysis in Condensed Matter}
\end{center}

\setcounter{equation}{0}
\setcounter{figure}{0}
\setcounter{table}{0}
\setcounter{page}{1}
\setcounter{section}{0}

\makeatletter

\renewcommand{\theequation}{S\arabic{equation}}
\renewcommand{\thesection}{S\arabic{section}}
\renewcommand{\thefigure}{S\arabic{figure}}

\section{Polyhedral Units identified in vicinal GBs using the PPM Algorithm}

\begin{figure}[h!]
  \centering
  \includegraphics[width=\linewidth]{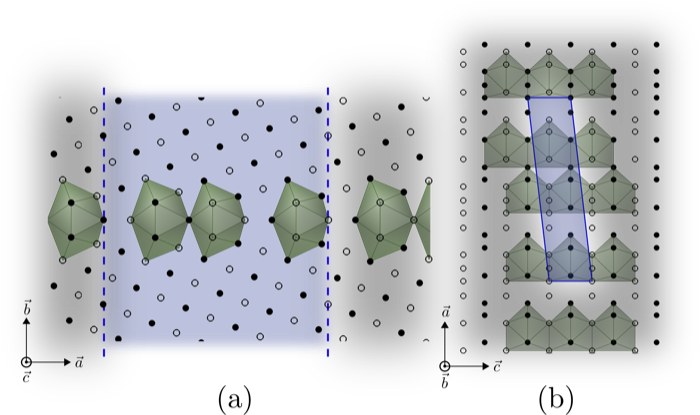}  
  \caption{The atomistic structure of $\Sigma 73 \left( 0 \, \bar{3} \, 8
    \right) $ GB is shown. The views in (a) and (b) are along the tilt axis and
    boundary-plane normal, respectively. The axes in the figure, $\vec{a}$,
    $\vec{b}$ and $\vec{c}$ correspond to the $[0 \, 8 \, 3]$, $[0 \, \bar{3} \,
    8]$ and $[1 \, 0 \, 0]$ lattice directions, respectively.}
  \label{fig:s73_gb1}
\end{figure}

\clearpage
\newpage

\begin{figure}[p]
  \centering
  \includegraphics[width=\linewidth]{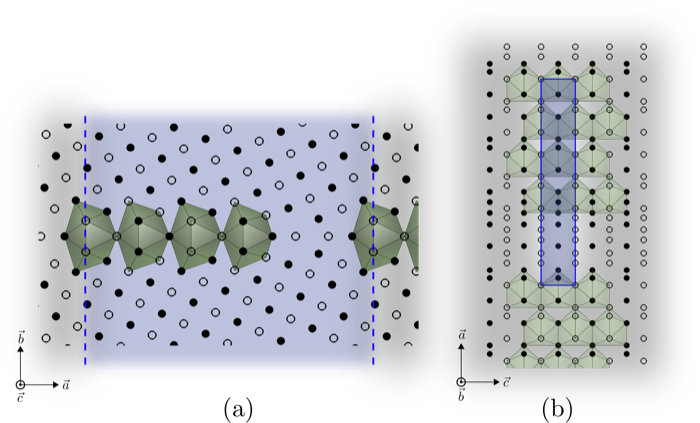}  
  \caption{The atomistic structure of $\Sigma 73 \left( 0 \, \bar{11} \, 5
    \right)$ GB is shown. The views in (a) and (b) are along the tilt axis and
    boundary-plane normal, respectively. The axes in the figure are such that,
    $\vec{a} = [0 \, \bar{5} \, \bar{11}]$, $\vec{b} = [0\, \bar{11}\, 5]$ and
    $\vec{c} = [\bar{1} 0 0]$.}
  \label{fig:s73_gb2}
\end{figure}

\clearpage
\newpage

\section{Structural Unit Model for the $[100]$ symmetric tilt GBs}

\begin{figure}[h!]
  \centering
  \includegraphics[width=0.75\linewidth]{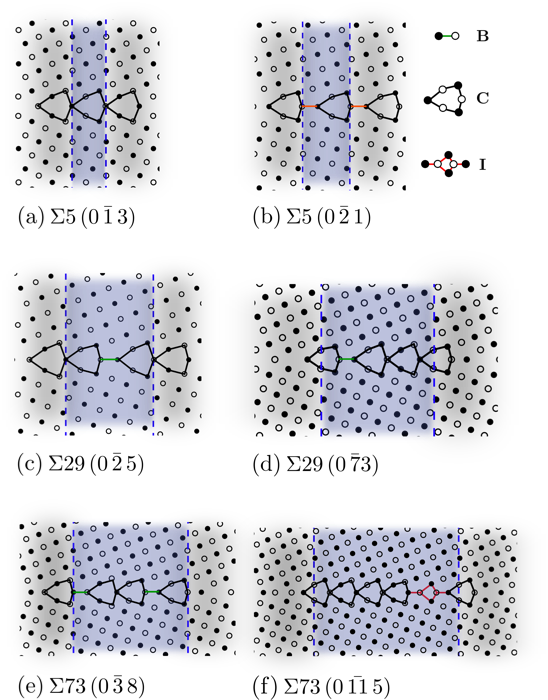}
  \caption{The Structural Unit Model representations of all the GBs analyzed in
    this article are shown.}
  \label{fig:sum}
\end{figure}